\documentclass[letterpaper]{sfchem}
\usepackage{graphicx}

\begin{document}

\title{Star Formation: 3D Collapse of Turbulent Cloud Cores}

\author{Michael A.\ Reid, Ralph E. Pudritz \& James Wadsley} 
  \institute{Department of Physics and Astronomy, McMaster University, 
  Hamilton, ON, Canada L8S 4R8}  
\authorrunning{Reid, Pudritz, and Wadsley}
\titlerunning{3D Collapse of Turbulent Cloud Cores}

\maketitle 

\begin{abstract}

We have performed fully 3-D simulations of the collapse of molecular cloud
cores which obey the logatropic equation of state.  By following the
collapse of these cores from states of near hydrodynamic equilibrium, we
are able to produce accretion histories which closely resemble those of
observed cores.  The accretion proceeds in four distinct stages: an
initial period of very slow accretion; a period of vigorous accretion
following the development of a central density singularity, with $\dot{M}
\propto t^{3}$, as predicted by self-similar models; a period of relatively
stable, vigorous accretion;  and finally a gradual decrease in the
accretion rate once about 50\% of the available mass of the molecular
cloud core has been accreted.  These results may explain the accretion
histories of cores as they pass through the pre-protostellar, Class 0, and
Class I stages. 


\end{abstract}

\section{Introduction}
	
	Mounting evidence indicates that massive molecular cloud cores
possess properties which are not adequately explained by models employing an
isothermal equation of state.  For example, surveys of prestellar cores in
a variety of star-forming complexes show that intermediate and high mass
cores have nonthermal line widths which are significantly greater than
their thermal line widths (\cite{cm95}).  Models of the collapse of 
isothermal molecular cloud cores exhibit a number of features which 
are not borne out by observations, such as a time-invariant rate of 
accretion onto the central object (\cite{Shu77}) and highly 
supersonic infall (\cite{fc93}).  

	To address some of the shortcomings of isothermal models,
\cite{mp96} advanced a model for molecular cloud core collapse which
accounts for some degree of turbulent support via a phenomenological
equation of state.  This logatropic equation of state reads \\
$P/P_{c}~=~1~+~A~\ln(\rho/\rho_{c})$ where the subscript ``c'' denotes
``central'' values.  The constant $A$ is an adjustable parameter whose
value for a range of molecular cloud core masses was determined
empirically by \cite*{mp96} to be $A = 0.2 \pm 0.02$. 

	The logatropic equation of state admits two equilibrium solutions. 
The features of each solution are developed in \cite*{mp96}.  The
``singular'' equilibrium solution is analogous to the singular isothermal
sphere (SIS) of \cite*{Shu77}, except that its density profile is $\rho
\propto r^{-1}$.  The ``nonsingular'' solution is analogous to the
Bonnor-Ebert sphere, in that it is a pressure-truncated equilibrium
solution with a finite central density and density profile in its outer 
regions which matches that of the singular solution.

	The nonsingular logatrope has the potential to be as useful a
model for massive star-forming cores as the Bonnor-Ebert sphere is for low
mass star-forming cores (see, for example, \cite*{alves}, who showed 
that a Bonnor-Ebert sphere is an excellent fit to the structure of 
dark cloud B68).  Hence, we undertook a numerical study of the
collapse of a nonsingular logatropic sphere from an initially
near-equilibrium state until most of its mass had been accreted onto the
central condensation. 

\section{Numerical Methods and Tests}

	Our fully 3D simulations were conducted using the ZEUS-MP
parallelized hydrodynamics code.  We used a 256$^{3}$ Cartesian grid with
a ``sink cell'' at its centre.  Mass falling into the sink cell was
presumed to have accreted onto the central protostar and was represented
thereafter by a point mass at the grid centre.  The initial configuration
for each of our simulations was a spherical molecular cloud core in near
hydrodynamic equilibrium, surrounded by an external medium of low and
constant pressure.  We conducted extensive tests of our numerical method
using the collapse of the singular logatrope (\cite{mp97}) as our test
case.  Excellent agreement between the analytic and numerical solutions
was found.  (see \cite*{rpw} for a full description of our numerical
method). 

\section{Results}

\begin{figure}
\resizebox{\hsize}{!}{\includegraphics{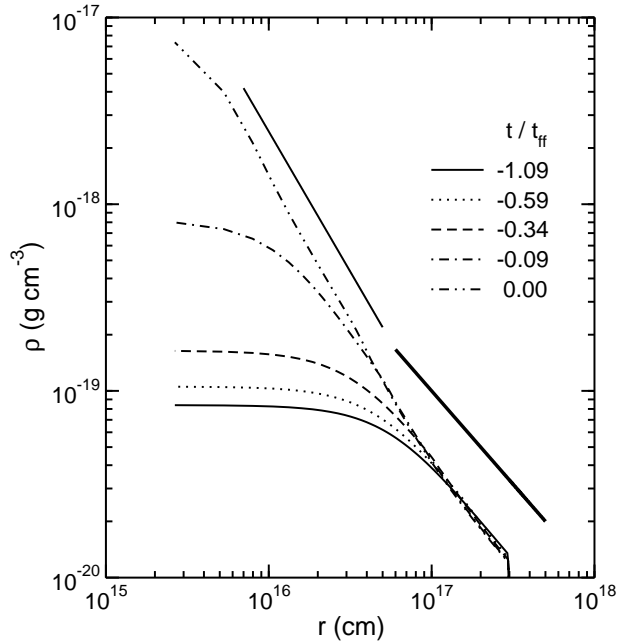}}
\caption{Density evolution of the 1~M$_{\odot}$ nonsingular core.  The 
singular density profile is reached at t=0.  The thick straight line 
indicates $\rho \propto r^{-1}$ and the thinner straight line indicates 
$\rho \propto r^{-3/2}$.  The solid curve ($t = -1.09 \bar{t}_{\rm ff}$) 
represents the initial density profile of the nonsingular core.  Adapted 
from Reid, Pudritz, \& Wadsley (2002).} \label{f9}
\end{figure}

\begin{figure}
\resizebox{\hsize}{!}{\includegraphics{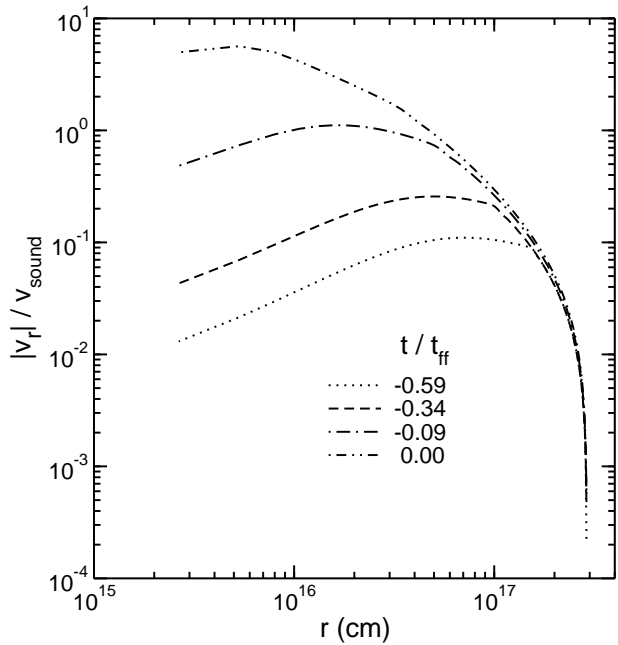}}
\caption{Evolution of the radial infall speed, $|v_{r}|$, of a 
1~M$_{\odot}$ nonsingular logatrope preceding the development of the 
singular density profile.  Adapted from Reid, Pudritz, \& Wadsley (2002).} 
\label{f10}
\end{figure}

	The most interesting of our results from the collapse of the
nonsingular logatrope are presented in Figures \ref{f9}-\ref{f11b}. 
Figures \ref{f9} and \ref{f10} show the evolution of the density and
infall speed profiles of a collapsing nonsingular logatrope prior to the
formation of a singular density profile.  The model has been scaled such
that the total mass of the core is 1 M$_{\odot}$.  The total core mass is
determined by the choice of central temperature and external
pressure---our choices of $T_{c} = 10$~K and $P_{s} = 1.3 \times
10^{5}$~k$_{B}$ result in a core of total mass 1~M$_{\odot}$, but the
results can be rescaled and applied to models of higher mass (for example,
by assuming a higher confining pressure).  Indeed, observational evidence
has emerged which indicates that the confining pressures on molecular
cloud cores are one or two orders of magnitude higher than we assumed
(\cite{john}). 

	Perhaps one of the most notable features of Figure \ref{f9} is the
behaviour of the density profile at late times.  At the moment when the
density becomes singular (labeled $t=0$), the density in the inner parts
of the core scales as $r^{-3/2}$.  This is the density profile expected
for a collapsing singular logatropic sphere, as well as for a collapsing
SIS.  This implies that, whereas observations of r$^{-3/2}$ density
profiles in molecular cloud cores have traditionally been interpreted as
evidence for isothermal collapse, this should no longer be the case.  Even
models which incorporate turbulent support will show such a density
profile. 

	Because of their extra nonthermal support, nonsingular logatropic
spheres collapse more ``gently'' than do their isothermal counterparts. 
 As can be seen by comparing our Figure \ref{f10} with Figure 1a of
\cite*{fc93}, the logatropic collapse is more gentle than that of the
Bonnor-Ebert sphere.  \cite*{fc93} find that, at the time of singularity
formation, 44\% of the mass of a Bonnor-Ebert sphere is in supersonic
infall.  We find that less than 5.5\% of a nonsingular logatrope is in
supersonic infall at the same point in its evolution.

	Figure \ref{f11b} shows the rate of accretion onto the central 
object in collapsing nonsingular logatropes of masses 1, 2.5, and 
5~M$_{\odot}$.  The reader should bear in mind that these are fiducial 
masses only---the models can be rescaled to produce much more massive 
cores and hence higher accretion rates by increasing the confining 
pressure.  Each curve shows the evolution of the accretion rate 
from the near-equilibrium starting point and until 95\%, 91\%, 
and 29\% of the total available mass has been accreted (for 
the 1, 2.5, and 5~M$_{\odot}$ cases respectively).  

\section{Discussion}

\begin{figure}
\resizebox{\hsize}{!}{\includegraphics{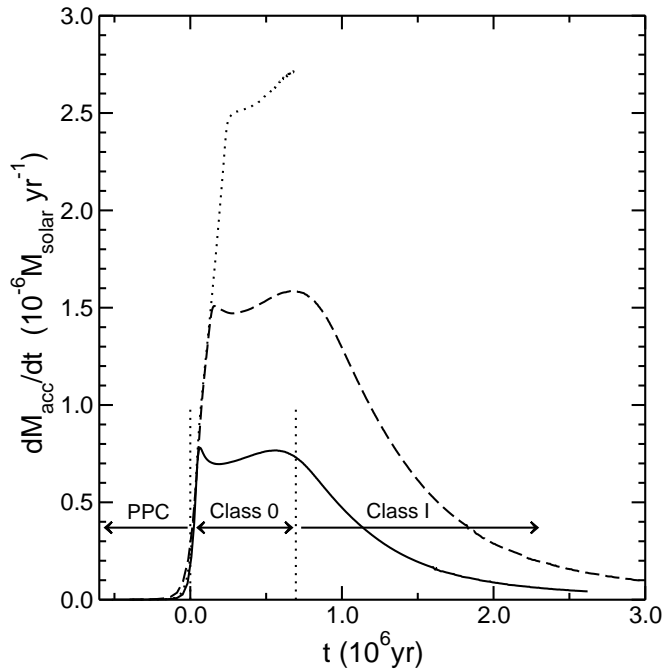}}
\caption{Time evolution of $M_{acc}(t)$ and $\dot{M}_{acc}(t)$ for the 
collapse of nonsingular logatropic spheres with dimensionless radii of 
$R/r_{o}$ = 1.34 (solid line), 2.21 (dashed line), and 3.26 (dotted 
line).  The data are scaled such that each core has a central 
temperature of 
$T_{c} = 10$~K and surface pressure of $P_{s} = 1.3 \times 
10^{5}$~k$_{B}$, giving them total masses of 1, 2.5, and 5~M$_{\odot}$.  The vertical dotted lines indicate suggested transitions 
between the PPC, Class O, and Class I stages for the 1~M$_{\odot}$ 
model.  Time $t=0$ marks the moment at which the core's density profile 
becomes singular.  Adapted from Reid, Pudritz, \& Wadsley (2002).} 
\label{f11b} 
\end{figure}

	The accretion profiles shown in Figure \ref{f11b} bear several
important similarities to what is known observationally about accretion in
young stellar objects.  A simple picture of the evolution of an accreting
molecular cloud core suggests an accretion history of the following form:
accretion proceeds slowly through the pre-protostellar (PPC) phase,
increases markedly at the formation of a central source (to power the
substantial outflows seen in Class O sources), and finally declines and
dissipates as the core passes into the Class I stage.  \cite*{andre} mark
the transition between the Class O and Class I phase at the point where
the mass accreted onto the central object is roughly equal to that
remaining in the envelope.  Our results show a similar transition at the
same point.  As indicated for the 1~M$_{\odot}$ curve in Figure
\ref{f11b}, there is an initial period of slow accretion, which we liken
to the PPC stage.  As soon as the density profile of the core becomes
singular ($t=0$), which can be thought of as the moment of protostar
formation, the accretion rate increases markedly, launching a phase 
of vigorous accretion sufficient to power an outflow.  We liken this to the
Class 0 phase.  Once half of the available mass has been accreted, the
accretion rate declines steadily to near-zero.

	We suggest that our simulations of the collapse of logatropic
molecular cloud cores bear significant similarities to the observations. 
They offer an alternative method for modelling the structure of
intermediate and higher mass molecular cloud cores.  Cores which exhibit
density power-law indices of between 1 and 1.5 (see, for example,
\cite*{vdt}) should be pursued as possible candidate nonsingular
logatropes. Our results bolster the argument that models of collapsing
protostellar cores \emph{must} account for some degree of nonthermal
support in order to reproduce the observations.

\begin{acknowledgements}
	
	We are grateful to the organizers of SFCHEM 2002 for allowing us 
to present our work.  The work of M. A. R. was supported in part by a 
scholarship from the Natural Sciences and Engineering Research Council of 
Canada (NSERC).  The work of R. E. P. was supported by NSERC.  J. W. is a 
SHARCNET Research Associate.

\end{acknowledgements}

\end{document}